\documentclass[pra, twocolumn,english,showpacs,floatfix]{revtex4}
\usepackage{graphics}
\usepackage{graphicx}
\usepackage{epsfig}
\usepackage{amssymb}
\usepackage{amsmath}

\begin{document}

\newcommand{\ket}[1]{|{#1}\rangle}
\newcommand{\bra}[1]{\langle{#1}|}
\newcommand{\braket}[1]{\langle{#1}\rangle}
\newcommand{\ad}{a^\dagger}
\newcommand{\e}{\ensuremath{\mathrm{e}}}
\newcommand{\norm}[1]{\ensuremath{| #1 |}}
\newcommand{\aver}[1]{\ensuremath{\big<#1 \big>}}
\renewcommand{\Im}{\operatorname{Im}}
\newcommand{\etal}{\textit{et al.~}}
\title{Modulation spectroscopy with ultracold fermions in an optical lattice}
\author{C. Kollath$^{1}$, A. Iucci$^{1}$, I. McCulloch$^{2}$,  T. Giamarchi$^{1}$}
\affiliation{$^1$ DPMC-MaNEP, University of Geneva, 24 Quai
Ernest-Ansermet, CH-1211 Geneva, Switzerland}
\affiliation{$^2$ Institute for Theoretical Physics, RWTH Aachen, D-52056 Aachen, Germany}

\begin{abstract}
We propose an experimental setup of ultracold
fermions in an optical lattice to determine the pairing gap in a
superfluid state and the spin ordering in a Mott-insulating state. The idea is to apply a periodic modulation of
the lattice potential and to use the thereby induced double occupancy to probe
the system. 
 We show by full time-dependent calculation using the adaptive time dependent
 density-matrix renormalization group method that the position of the peak
 in the spectrum of the
 induced double occupancy gives the pairing energy in a superfluid and the
 interaction energy in a Mott-insulator, respectively. In the Mott-insulator
 we relate the spectral weight of the peak to the spin ordering at finite
 temperature using
 perturbative calculations.  
\end{abstract}
\maketitle

The use of Feshbach resonances \cite{RegalJin2004} and the loading of ultracold
atoms into optical lattices \cite{OrzelKasevich2001, GreinerBloch2002}
opened up the possibility to realize strong interactions in quantum
gases.
These systems provide outstanding control over most of their
parameters, which offers the possibility to use them as `quantum simulators' \cite{Feynman1982}, i.e.~to
investigate unresolved questions from quantum many-body physics in a
well-controllable system.
One example is the crossover between a Bose-Einstein condensate of
bosonic molecules
and a BCS-like state of fermionic atoms \cite{GreinerJin2005,ChinGrimm2004}.
A major step in the direction to `simulate' phenomena in periodic potentials has
been taken by K\"ohl \etal\cite{KoehlEsslinger2005} and very recently by Chin
\etal\cite{ChinKetterle2006} who succeeded in
loading ultracold fermions into a three-dimensional optical lattice.

However, the
possibilities to probe ultracold gases in optical lattices are still very
limited as compared to measurement techniques in condensed matter systems.
On top of this some of the known probes for bosonic gases are not useful in
the context of fermionic gases. Thus the lately experimental realization demands for new techniques.
We propose an experimental setup to measure the pairing energy of a
superfluid state and the interaction strength in a Mott-insulating or
liquid state of a two component mixture
of fermionic atoms in an optical lattice. Proposals how to gain information
about the pairing in a superfluid state have been
made for example in
Ref.~\cite{Carusotto2005,BuechlerZwerger2004,AltmanLukin2003} for a system
without a lattice and in Ref.~\cite{HofstetterLukin2002} with a lattice.
The setup proposed here is based on the experiment by St\"oferle
\etal\cite{StoeferleEsslinger2004} in which a gas of ultracold bosons was
probed using the energy absorption induced by periodic modulations of the
lattice potential.
It has been shown that very precise information about the
excitation structure of a bosonic system
\cite{vanOostenStoof2005,KraemerDalfovo2005,IucciGiamarchi2005,ReischlUhrig2005,PupilloBatrouni2006,KollathSchollwoeck2006},
and about the degree of incommensurate filling \cite{KollathSchollwoeck2006}
can be extracted by means of this experiment. It would therefore be desirable
to extent the measurement to fermionic systems. 
However, in a fermionic system the energy absorption cannot be determined in the same way as it was
done in the bosonic setup. There the energy absorption imprints a clear characteristic
signal in time-of-flight measurements which is not the case for
fermionic systems.
In the present work we therefore propose to probe a two-component fermionic
mixture by the observation of the double
occupancy which is induced by a periodic modulation of the lattice potential. The double occupancy can be observed by projecting double occupied sites onto molecules
\cite{StoeferleEsslinger2006}.
We show that this measurement gives information on the pair binding energy
of the superfluid state and the interaction energy in a
liquid or Mott-insulating state. We further propose how information on the spin
ordering could be extracted in the case of repulsive interaction.

In our studies we focus on a two component fermionic gas which
is tightly confined to one dimensional tubes along the $x$-direction by a strong lattice in the
transverse directions. Such a system was realized in the
experiment \cite{MoritzEsslinger2005}. If an additional periodic potential
$V_x$ is applied along the
tubes, the system can be
described \cite{JakschZoller1998, HofstetterLukin2002} by the Hubbard model
\begin{equation}
\label{eq:bh}
H= -J \sum_{j,\sigma} \left(c_{j,\sigma}^\dagger
c^{\phantom{\dagger}}_{j+1,\sigma}+h.c.\right)  + U \sum _{j,\sigma}
\hat{n}_{j,\uparrow} \hat{n}_{j,\downarrow}.
\end{equation}
Here $\sigma=\{ \uparrow , \downarrow \}$ labels the two hyperfine states,
$c^\dagger_{j,\sigma}$ and $c_{j,\sigma}$ are the corresponding
creation and annihilation operators, and $ \hat{n}_{j,\sigma}=
c^\dagger_{j,\sigma} c^{\phantom{\dagger}}_{j,\sigma}$ is the number operator
on site $j$. The values of hopping parameter $J$ and the interaction strength $U$
depend among other experimental parameters on the lattice
potential $V_x$ \footnote{In our simulation we use 
parameters corresponding to a gas of $\textrm{K}^{40}$
atoms in the $F=9/2$, $m_f=-9/2$ and $m_f=-7/2$
states and $\lambda=826$nm.}.  
The equilibrium properties of the one-dimensional Hubbard model have been intensively studied by a variety of
techniques \cite{Gebhard1997,EsslerKorepin2005,Schulz1995,Giamarchibook},
e.g. exact solutions, bosonization, and numerical methods.
At low temperature in a system with attractive interaction a
superfluid state is formed, whereas for a repulsive interaction the
ground state is either a half-filled Mott insulator or, away from
half-filling, a Luttinger liquid.
However, the non-equilibrium properties are still a challenge. We use the quasi-exact
adaptive time-dependent density matrix renormalization group method (adaptive t-DMRG)
\cite{DaleyVidal2004,WhiteFeiguin2004, KollathZwerger2005} and perturbative calculations to
determine the time-evolution of the system.
A periodic modulation of the lattice potential along the tube 
$V_x(t)=V_0  \left(1+\delta V \cos{(\omega t)}\right) $, where $\omega$ is the
modulation frequency, can be
translated in the lattice description to a periodic
modulation of $J$ and $U$, i.e.~$J[V_x(t)]$ and
$U[V_x(t)]$. Thus the system is described by an explicitely time-dependent
many-body Hamiltonian. In the following $J_0$ and $U_0$ will be used for the
values $J[V_0]$ and $U[V_0]$.

A system subjected to a periodic modulation of the lattice potential will
on average absorb energy if the
modulation frequency $\omega$ corresponds approximately to the energy $\Delta E$
needed to create an excitation, i.e.~$\hbar \omega \approx \Delta E$.
In the limit of strongly attractive interaction one important class of
excitations is the breaking of pairs. In the strongly repulsive limit particle-hole
excitations are formed. In both cases these excitations cost an
energy of the order of the interaction energy $U_0$.
Therefore we expect that the application of a periodic lattice modulation can cause energy absorption, if the frequency corresponds to
the excitation energy, i.e.~$\hbar \omega \approx U_0$. The two cases of attractive and repulsive interaction are
related by an electron-hole transformation for one spin species, e.g. $c_{\downarrow, j} \rightarrow
(-1)^j c^\dagger_{\downarrow, j}$, which maps $U \rightarrow -U$
\cite{Gebhard1997,Giamarchibook}. Thus one can restrict the parameters to one sign of the
interaction only. 

\begin{figure} [ht]
  \begin{center}
    {\epsfig{figure=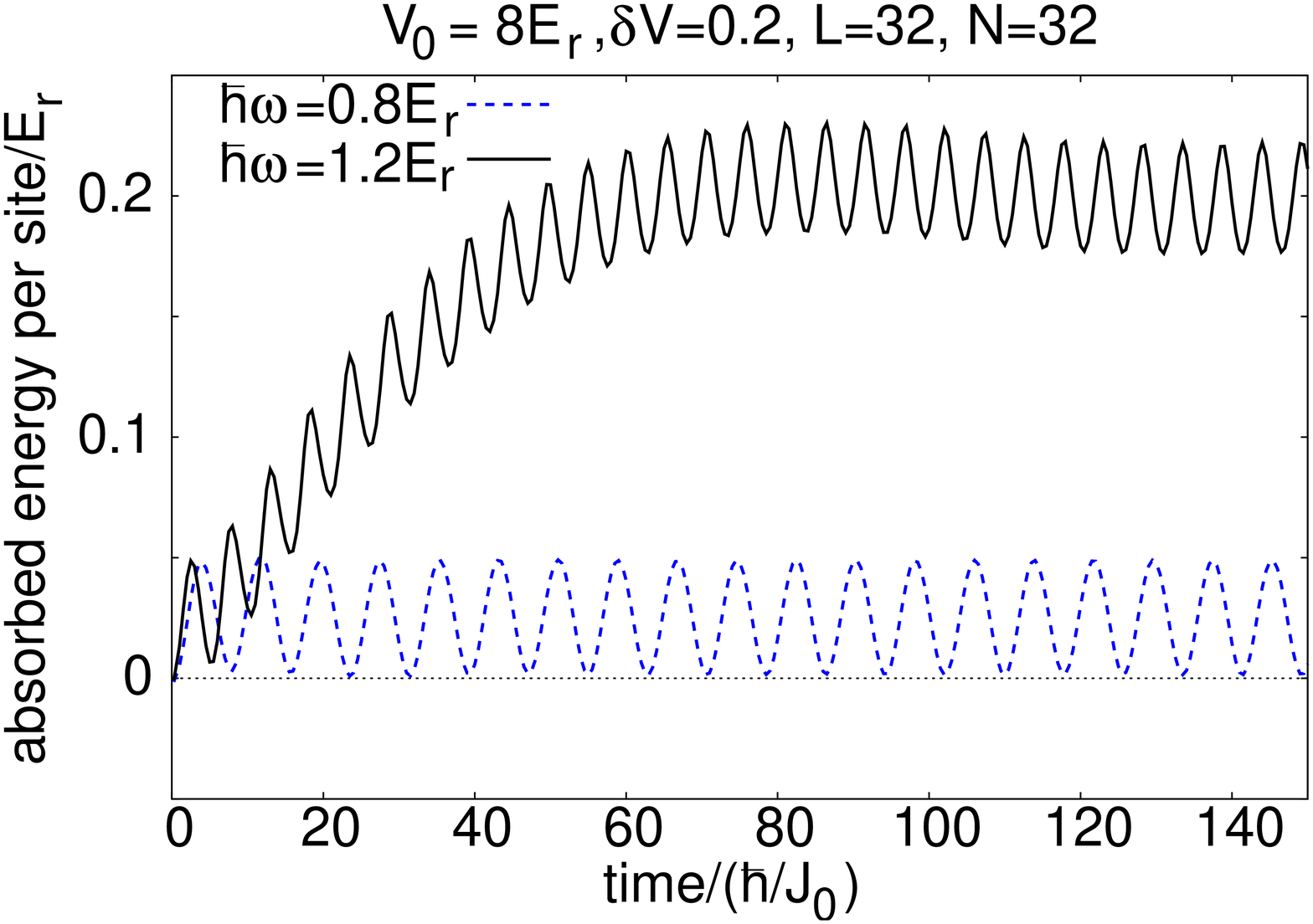,width=0.7\linewidth}}
  \end{center}
  \caption{Absorbed energy $E(t)$ versus time for different modulation
  frequencies for an initial lattice potential height of $V_0$. If the modulation frequency
  is close to resonance ($\hbar \omega = 1.2 E_r$), energy is absorbed, whereas
  off resonance ($\hbar \omega = 0.8 E_r$) no energy
  absorption takes place.
  }
  \label{fig:energy}
\end{figure}

In Fig. \ref{fig:energy}
the energy absorption for the case of a periodic modulation, both close
to resonance $\hbar \omega \approx U_0$ and off-resonance is shown. In both
cases the
absorbed energy shows an oscillatory behaviour with the frequency
corresponding to the modulation frequency. Off-resonance on average no energy is
absorbed by the system, whereas at resonance a clear absorption of energy
is visible which saturates at longer times.

\begin{figure} [ht]
  \begin{center}
    {\epsfig{figure=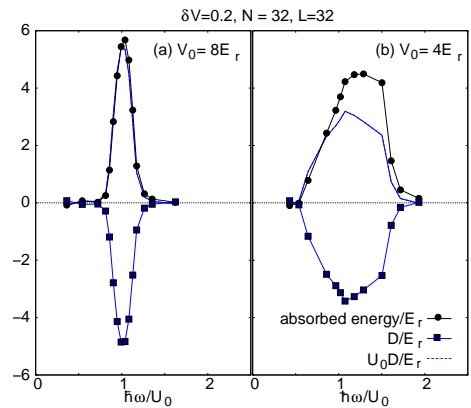,width=0.7\linewidth}}
  \end{center}
  \caption{Energy absorption spectrum (circles) and induced double occupancy $D$
  (squares) for (a) an initial lattice potential
  $V_0=8E_r$ and (b) an initial lattice potential
  $V_0=4E_r$ for a strong modulation $\delta V=0.02$ and $a_s=-9.2$nm. 
A clear peak around $\hbar \omega \approx U_0$ arises in both quantities. For
  a better comparison $U_0 D$ is shown as well.
  }
  \label{fig:energyspectrum}
\end{figure}
The energy absorption versus the modulation frequency is shown in
Fig.~\ref{fig:energyspectrum}.  
Here the average value at $t_m=50\hbar /E_r$ is
taken as a measure
for the energy absorption \footnote{This is determined by smoothening the oscillations in the curve by convolution with a Gaussian
$\exp{\left( -t^2/(2\Delta T^2)\right)}$ with $\Delta T \approx
  \pi/\omega$.}. 
For an initial potential $V_0=8E_r$ a clear peak centered around the
resonance frequency $\hbar \omega \approx U_0$ is
seen (Fig.~\ref{fig:energyspectrum} (a)). For a lower initial potential $V_0=4E_r$
(Fig.~\ref{fig:energyspectrum} (b)) we see a broadening of the peak and a
clear asymmetric form. The maximum of the peak is shifted slightly above the
value $\hbar \omega \approx U_0$. These changes are due to the fact that the distance of
the ground state energy to the center of the next band is not exactly given by
the energy $U_0$ but shows corrections in
$J_0$. 

In the experiment with bosonic atoms \cite{StoeferleEsslinger2004} the
broadening of the characteristic low momentum peak in the
time-of-flight images has been used as a measure for the absorbed
energy.
In the fermionic experiment the observation of the absorbed energy is more
difficult. Due to Pauli's principle for the momentum distribution all momenta up to the Fermi
momentum are occupied. Therefore the absorption of energy will only cause a smearing out
of the step around the Fermi momentum in time-of-flight images which can not be used to determine
quantitatively the
amount of energy absorbed \footnote{A similar smearing is caused
  by the finite temperature and the trapping potential \cite{RigolMuramatsu2004c}}.
We propose to use the double occupancy in the lattice, i.e.~
$D=\aver{\sum_j n_{j,\uparrow} n_{j,\downarrow}}$ as a quantity
which can be observed more easily by projecting pairs of atoms onto molecules
\cite{StoeferleEsslinger2006}. Since for strongly attractive
(repulsive) interactions the energy absorption corresponds to a
breaking of pairs (generation of particle-hole excitations) we
expect that in this limit the energy absorption shows up in
the expectation value of the double occupancy.
In perturbative calculations we find that up to
the quadratic response regime the average induced double
occupancy and the average absorbed energy are related simply by a
factor $U_0$. This result is obtained by expanding $J(t)$ up to first
order in $\delta V$, neglecting the time-dependence in the interaction,
i.e. $U(t)\equiv U_0$, and writing the Hamiltonian as $H(t)=H_0 +
g(t)H_K$, where $H_0$ is the Hubbard Hamiltonian at $t=0$,
$H_K$ is the hopping operator, and
$g(t)=(dJ/dV_0)_{\delta V=0}\delta V \cos(\omega t)$. The linear response to this time dependent
perturbation gives a purely oscillating contribution to the double
occupancy, which averages to zero. Therefore, we consider the
quadratic response, which is given by

\begin{equation}
D_2\left(t\right)  =\frac{1}{U_0}\int_{-\infty}^{\infty}\int_{-\infty}^{\infty}%
dt^{\prime}dt^{\prime\prime}g(t') g(t'')\chi_2\left(
t,t^{\prime},t^{\prime\prime}\right)
\end{equation}
where $\chi_2\left(t,t^{\prime},t^{\prime\prime}\right)$ is the
quadratic susceptibility. Its relevant term, i.e.~the one that
gives a non trivial time dependence, can be related to the first
order susceptibility,
\begin{equation}
\chi^\mathrm{rel}_2\left(t,t^{\prime},t^{\prime\prime}\right)=-\theta(t'-t'')\frac{d}{dt}\chi_K(t-t')+\mathrm{osc.\;terms},
\end{equation}
where $\theta(t)$ is the step function, and
$\chi_K(t)=(1/i\hbar)\left\langle\left[H_K(t),H_K(0)\right]\right\rangle$.
The final expression for the double occupancy is
$D_2(t)=D_2(0)-\frac{1}{2}[g(0)]^2\omega t\Im\tilde{\chi}_K(\omega)/U_0+\mathrm{osc.\;terms}$, where $\tilde{\chi}_K(\omega)$ is the Fourier transform
of $\chi_K(t)$. One finds the same linear growth with time for the
expectation value of the total energy, without the factor $1/U_0$. The oscillatory terms, however, are
different for both quantities.

In the experimental realization a relatively strong modulation strength $\delta
V$ should
be used to get an observable signal. We checked numerically that even
for modulation strengths of $20\%$, the structure of the energy absorption spectrum and the spectrum
of the induced double occupancy still contain similar information.
In particular we find that the position of the peak agrees
for both, whereas deviations occur in the width and amplitude for
higher values of the perturbation. Examples for the comparison are shown in Fig. \ref{fig:energyspectrum}.

Since we have seen that the experimentally measurable quantity, the induced
double occupancy, can be used as a similar measure as the absorbed energy, we
will discuss in the following what kind of information can be
extracted from
this quantity. Hereby we focus on (i) how to obtain a good measure of the
energy necessary to break a pair in the superfluid state or to create a
particle-hole excitation in a Mott-insulator or liquid state and
(ii) how to gain information about the spin-ordering in a state.

\begin{figure} [ht]
  \begin{center}
{\epsfig{figure=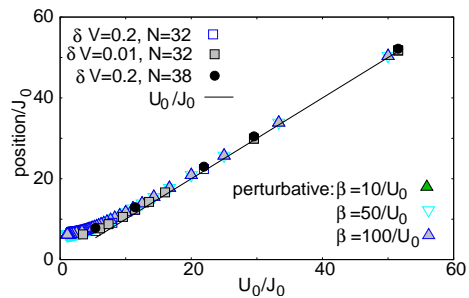,width=0.7\linewidth}}
  \end{center}
  \caption{ The position of the peak in the induced double occupancy.
Results obtained by a full time-dependent calculation (adaptive t-DMRG) are
  shown at half filling, $a_s=9.2$nm for a strong
  perturbation $\delta V=0.2$ (empty squares), at half filling for a weak
  perturbation $\delta V=0.01$ (filled squares), and away
  of half filling for a strong perturbation $\delta V=0.2$
  (circles). Perturbative results (triangles) are shown for different temperatures $k_B
  T=1/\beta$ for small systems sizes ($L=6$ and $N=6$). 
 A good
  agreement between all results is found, which makes the
  position of the peak a good measure of $U_0/J_0$.
  }
  \label{fig:position}
\end{figure}
 Direct access to the average energy needed to break
pairs or to create a particle hole excitations can be obtained from the
position of the absorption peak. At zero temperature  the dependence of the
peak position on the initial interaction strength $U_0/J_0$ is shown in
Fig.~\ref{fig:position} for different modulation strengths and fillings (marked by squares and
cirles). The
results were obtained performing the full time-evolution of the system using the adaptive t-DMRG \footnote{We determine the position $a$ by
fitting a Gaussian of the form $c \exp\left(-(x-a)^2/\sigma^2\right)$.
Note, that the fit by a Gaussian function is not very accurate for small
values of $U_0/J_0$ due to the asymmetric form of the peak. }.
The position of the peak depends only on the initial interaction strength and not on the size of
the perturbation. For strong interactions the dependence is linear.
For small values of the interaction strength $U_0/J_0$ it deviates
from the linear form, since the delocalization of the fermions becomes more
important and the simple picture of assuming the state to consist of
localized particles breaks down \cite{Gebhard1997,EsslerKorepin2005}.
To study the robustness of the peak to the influences of finite
temperature we
additionally perform a perturbative calculation for small system sizes. In
Fig.~\ref{fig:position} we show the position of the maximum value of the
contributions in quadratic response calculations \footnote{We show the position of the maximum value of the
  $\delta$-peaks, i.e. $1/Z (E_m-E_n) \norm{\langle n|H_K|
   m \rangle}^2 \exp (-\beta (E_n-E_m) (1-\exp (-\beta
  (E_m-E_n)))$. Here $E_n$ are the eigenvalues of the unperturbed Hamiltonian,
$\beta$ gives the inverse temperature.}. The collapse
of the data for different temperatures shows that the position is mainly
independent of temperature. The good agreement between our results shows that the position is a very robust quantity,
which is almost independet of temperature, strength of the modulation, size and filling of the system.

In experiments an additional harmonic trapping potential is present. We expect
that its main influence on the induced double occupancy is a broadening of
the width of the peak, as was found in the
bosonic case \cite{KollathSchollwoeck2006}. In particular, the position of the
peak will remain unaffected.
In the limit of strong interaction this can be understood by the introduction of potential energy
differences between neighbouring sites by the trapping potential. This causes
a shift of the excitation energy for the pair breaking or the creation
of particle-hole pairs, but with a sign depending on the direction of the
excitation. Therefore the position of the peak remains unchanged whereas its
form can be broadened. 
In the experiment with bosonic atoms \cite{StoeferleEsslinger2004} for strong interaction strength
the position of the resonance peak in the energy absorption could
be determined approximately up to an accuracy of 20 Hz. Note, that since the
broadening of the peak is proportional to $J_0$ the accuracy is best for strong interaction. In the fermionic case
it is hard to estimate the loss of accuracy by the projection onto molecules, but
we expect that the accuracy which can be obtained is of the same order of
magnitude. 

 \begin{figure} [ht]
  \begin{center}
   {\epsfig{figure=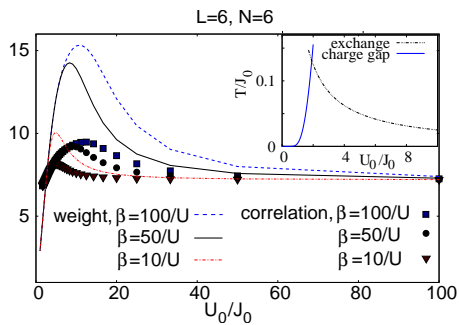,width=0.7\linewidth}}
  \end{center}
  \caption{The weight of the peak in the induced double occupancy for different
  temperatures $1/\beta$ compared to the nearest neighbour correlations
  function  $ 4L\aver{n_{\uparrow,j}n_{\downarrow,j+1}}$ in a small
  system with $L=6$ sites and $N=6$ particles. The inset shows the competing
  energy scales influencing the spin ordering.
  }
  \label{fig:ampl}
\end{figure}
In a two component fermionic system with repulsive interactions the probability
to create a particle-hole excitation depends
strongly on the spin ordering present in the intitial state. In a
ferromagnetically ordered state with only a few domain walls particle-hole
excitations are strongly supressed by the Pauli principle, whereas in an antiferromagnetically ordered state the
creation of particle hole pairs is facilitated. This means that a state which is
antiferromagnetically ordered will show more absorption than a paramagnetically or
ferromagnetically ordered state.  Thus the weight of the absorption peak
gives information about the spin ordering. To be more specific, for large $U_0/J_0$
perturbative calculations yield that the weight is proportional to the nearest
neighbour correlation $\aver{n_{\uparrow,j}n_{\downarrow,j+1}}$. 
This can be seen in Fig.~\ref{fig:ampl} where we compare perturbation results
for the weight of the absorption and nearest neighbour correlations for
different temperatures. At large values of 
$U_0/J_0$ we find very good agreement between the quantities, whereas for
smaller interaction the values deviate. 
A maximum occurs in both quantities at approximately the same finite
interaction strength. This is due to two competing
processes which destroy the spin order, the delocalization of charge
dominating at low $U_0/J_0$ and the
spin exchange dominating at large $U_0/J_0$. In the inset in
Fig.~\ref{fig:ampl} this competition of the energy scales for the spin
ordering, the charge gap and the spin exchange coupling are shown.
The maximum in the weight shows the crossover between the two energy
scales thus giving information about the spin ordering of the system. 

To summarize we have proposed a setup to measure the pair-binding energy, the interaction strength, and the spin-ordering in a Fermi gas in an optical lattice. The proposed
experiment observes the double occupancy induced by periodic lattice
modulations via the creation of molecules. The robustness of the
measurement procedure to external
influences such as temperature or a harmonic trapping potential makes an
precise experimental determination of the pair-binding energy, the interaction strength -- thereby the
scattering length -- and the spin ordering in the optical lattice possible. 
Since for strong interactions the proposed measurement is based on a local effect,
i.e. the creation of particle-hole pairs, we expect it to
hold as well in higher dimensions.

We would like to thank M.~K\"ohl, A.J.~Millis, A.~Muramatsu, and A.~Schadschneider for fruitful discussions. This
work was partly supported by the Swiss National Fund under MaNEP and
Division II.



\begin{thebibliography}{30}
\expandafter\ifx\csname natexlab\endcsname\relax\def\natexlab#1{#1}\fi
\expandafter\ifx\csname bibnamefont\endcsname\relax
  \def\bibnamefont#1{#1}\fi
\expandafter\ifx\csname bibfnamefont\endcsname\relax
  \def\bibfnamefont#1{#1}\fi
\expandafter\ifx\csname citenamefont\endcsname\relax
  \def\citenamefont#1{#1}\fi
\expandafter\ifx\csname url\endcsname\relax
  \def\url#1{\texttt{#1}}\fi
\expandafter\ifx\csname urlprefix\endcsname\relax\def\urlprefix{URL }\fi
\providecommand{\bibinfo}[2]{#2}
\providecommand{\eprint}[2][]{\url{#2}}

\bibitem[{\citenamefont{Regal et~al.}(2003)\citenamefont{Regal, Greiner, and
  Jin}}]{RegalJin2004}
\bibinfo{author}{\bibfnamefont{C.~A.} \bibnamefont{Regal}},
  \bibinfo{author}{\bibfnamefont{M.}~\bibnamefont{Greiner}}, \bibnamefont{and}
  \bibinfo{author}{\bibfnamefont{D.~S.} \bibnamefont{Jin}},
  \bibinfo{journal}{Phys.~ Rev.~ Lett.} \textbf{\bibinfo{volume}{92}},
  \bibinfo{pages}{040403} (\bibinfo{year}{2003}).

\bibitem[{\citenamefont{Orzel et~al.}(2001)\citenamefont{Orzel, Tuchman,
  Fenselau, Yasuda, and Kasevich}}]{OrzelKasevich2001}
\bibinfo{author}{\bibfnamefont{C.}~\bibnamefont{Orzel} {\em et al.}},
  \bibinfo{journal}{Science} \textbf{\bibinfo{volume}{291}},
  \bibinfo{pages}{2386} (\bibinfo{year}{2001}).

\bibitem[{\citenamefont{Greiner et~al.}(2002)\citenamefont{Greiner, Mandel,
  Esslinger, H\"ansch, and Bloch}}]{GreinerBloch2002}
\bibinfo{author}{\bibfnamefont{M.}~\bibnamefont{Greiner {\em et al.}}},
  \bibinfo{journal}{Nature} \textbf{\bibinfo{volume}{415}}, \bibinfo{pages}{39}
  (\bibinfo{year}{2002}).

\bibitem[{\citenamefont{Feynman}(1982)}]{Feynman1982}
\bibinfo{author}{\bibfnamefont{R.}~\bibnamefont{Feynman}},
  \bibinfo{journal}{Int.~J.~of theor.~ Phys.~} \textbf{\bibinfo{volume}{21}}, \bibinfo{pages}{467} (\bibinfo{year}{1982}).

\bibitem[{\citenamefont{Greiner et~al.}(2005)\citenamefont{Greiner, Regal, and
  Jin}}]{GreinerJin2005}
\bibinfo{author}{\bibfnamefont{M.}~\bibnamefont{Greiner}},
  \bibinfo{author}{\bibfnamefont{C.}~\bibnamefont{Regal}}, \bibnamefont{and}
  \bibinfo{author}{\bibfnamefont{D.}~\bibnamefont{Jin}},
  \bibinfo{journal}{Phys.~ Rev.~ Lett.} \textbf{\bibinfo{volume}{94}},
  \bibinfo{pages}{070403} (\bibinfo{year}{2005}).

\bibitem[{\citenamefont{Chin et~al.}(2004)\citenamefont{Chin, Bartenstein,
  Altmeyer, Riedl, Jochim, Hecker-Denschlag, and Grimm}}]{ChinGrimm2004}
\bibinfo{author}{\bibfnamefont{C.}~\bibnamefont{Chin {\em et al.}}},
  \bibinfo{journal}{Science} \textbf{\bibinfo{volume}{305}},
  \bibinfo{pages}{1128} (\bibinfo{year}{2004}).

\bibitem[{\citenamefont{K\"ohl et~al.}(2005)\citenamefont{K\"ohl, Moritz,
  St\"oferle, G\"unter, and Esslinger}}]{KoehlEsslinger2005}
\bibinfo{author}{\bibfnamefont{M.}~\bibnamefont{K\"ohl {\em et al.}}},
  \bibinfo{journal}{Phys.~ Rev.~ Lett.} \textbf{\bibinfo{volume}{94}},
  \bibinfo{pages}{080403} (\bibinfo{year}{2005}).

\bibitem[{\citenamefont{Chin et~al.}(2006)\citenamefont{Chin, Miller, Liu,
  Starn, Setiawan, Sanner, Xu, and Ketterle}}]{ChinKetterle2006}
\bibinfo{author}{\bibfnamefont{J.}~\bibnamefont{Chin {\em et al.}}},
  \bibinfo{journal}{cond-mat/0607004}  (\bibinfo{year}{2006}).

\bibitem[{\citenamefont{Carusotto}(2005)}]{Carusotto2005}
\bibinfo{author}{\bibfnamefont{I.}~\bibnamefont{Carusotto}},
  \bibinfo{journal}{cond-mat/0512539}  (\bibinfo{year}{2005}).

\bibitem[{\citenamefont{B\"uchler et~al.}(2004)\citenamefont{B\"uchler, Zoller,
  and Zwerger}}]{BuechlerZwerger2004}
\bibinfo{author}{\bibfnamefont{H.~P.} \bibnamefont{B\"uchler}},
  \bibinfo{author}{\bibfnamefont{P.}~\bibnamefont{Zoller}}, \bibnamefont{and}
  \bibinfo{author}{\bibfnamefont{W.}~\bibnamefont{Zwerger}},
  \bibinfo{journal}{Phys.~ Rev.~ Lett.} \textbf{\bibinfo{volume}{93}},
  \bibinfo{pages}{080401} (\bibinfo{year}{2004}).

\bibitem[{\citenamefont{Altman et~al.}(2004)\citenamefont{Altman, Demler, and
  Lukin}}]{AltmanLukin2003}
\bibinfo{author}{\bibfnamefont{E.}~\bibnamefont{Altman}},
  \bibinfo{author}{\bibfnamefont{E.}~\bibnamefont{Demler}}, \bibnamefont{and}
  \bibinfo{author}{\bibfnamefont{M.~D.} \bibnamefont{Lukin}},
  \bibinfo{journal}{Phys.~ Rev.~ A} \textbf{\bibinfo{volume}{70}},
  \bibinfo{pages}{013603} (\bibinfo{year}{2004}).

\bibitem[{\citenamefont{Hofstetter et~al.}(2002)\citenamefont{Hofstetter,
  Cirac, Zoller, Demler, and Lukin}}]{HofstetterLukin2002}
\bibinfo{author}{\bibfnamefont{W.}~\bibnamefont{Hofstetter {\em et al.}}},
  \bibinfo{journal}{Phys.~ Rev.~ Lett.}  (\bibinfo{year}{2002}).

\bibitem[{\citenamefont{St\"oferle et~al.}(2004)\citenamefont{St\"oferle,
  Moritz, Schori, K\"ohl, and Esslinger}}]{StoeferleEsslinger2004}
\bibinfo{author}{\bibfnamefont{T.}~\bibnamefont{St\"oferle {\em et al.}}},
  \bibinfo{journal}{Phys.~ Rev.~ Lett.} \textbf{\bibinfo{volume}{92}},
  \bibinfo{pages}{130403} (\bibinfo{year}{2004}).

\bibitem[{\citenamefont{van Oosten et~al.}(2005)\citenamefont{van Oosten,
  Dickerscheid, Farid, van~der Straten, and Stoof}}]{vanOostenStoof2005}
\bibinfo{author}{\bibfnamefont{D.}~\bibnamefont{van Oosten {\em et al.}}},
\bibinfo{journal}{Phys.~ Rev.~ A}
  \textbf{\bibinfo{volume}{71}}, \bibinfo{pages}{021601(R)}
  (\bibinfo{year}{2005}).

\bibitem[{\citenamefont{Kr\"amer et~al.}(2005)\citenamefont{Kr\"amer, Tozzo,
  and Dalfovo}}]{KraemerDalfovo2005}
\bibinfo{author}{\bibfnamefont{M.}~\bibnamefont{Kr\"amer}},
  \bibinfo{author}{\bibfnamefont{C.}~\bibnamefont{Tozzo}}, \bibnamefont{and}
  \bibinfo{author}{\bibfnamefont{F.}~\bibnamefont{Dalfovo}},
  \bibinfo{journal}{Phys.~ Rev.~ A} \textbf{\bibinfo{volume}{71}},
  \bibinfo{pages}{061602(R)} (\bibinfo{year}{2005}).

\bibitem[{\citenamefont{Iucci et~al.}(2005)\citenamefont{Iucci, Cazalilla, Ho,
  and Giamarchi}}]{IucciGiamarchi2005}
\bibinfo{author}{\bibfnamefont{A.}~\bibnamefont{Iucci}},
  \bibinfo{author}{\bibfnamefont{M.}~\bibnamefont{Cazalilla}},
  \bibinfo{author}{\bibfnamefont{A.}~\bibnamefont{Ho}}, \bibnamefont{and}
  \bibinfo{author}{\bibfnamefont{T.}~\bibnamefont{Giamarchi}},
  \bibinfo{journal}{cond-mat}  (\bibinfo{year}{2005}).

\bibitem[{\citenamefont{Reischl et~al.}(2005)\citenamefont{Reischl, Schmidt,
  and Uhrig}}]{ReischlUhrig2005}
\bibinfo{author}{\bibfnamefont{A.}~\bibnamefont{Reischl}},
  \bibinfo{author}{\bibfnamefont{K.}~\bibnamefont{Schmidt}}, \bibnamefont{and}
  \bibinfo{author}{\bibfnamefont{G.}~\bibnamefont{Uhrig}},
  \bibinfo{journal}{Phys.~ Rev.~ A} \textbf{\bibinfo{volume}{72}},
  \bibinfo{pages}{063609} (\bibinfo{year}{2005}).

\bibitem[{\citenamefont{Pupillo et~al.}(2006)\citenamefont{Pupillo, Rey, and
  Batrouni}}]{PupilloBatrouni2006}
\bibinfo{author}{\bibfnamefont{G.}~\bibnamefont{Pupillo}},
  \bibinfo{author}{\bibfnamefont{A.~M.} \bibnamefont{Rey}}, \bibnamefont{and}
  \bibinfo{author}{\bibfnamefont{G.~G.} \bibnamefont{Batrouni}},
  \bibinfo{journal}{cond-mat/0602240}  (\bibinfo{year}{2006}).

\bibitem[{\citenamefont{Kollath et~al.}(2006)\citenamefont{Kollath, Iucci,
  Giamarchi, Hofstetter, and Schollw\"ock}}]{KollathSchollwoeck2006}
\bibinfo{author}{\bibfnamefont{C.}~\bibnamefont{Kollath {\em et al.}}},
  \bibinfo{journal}{Phys.~ Rev.~ Lett.} \textbf{\bibinfo{volume}{97}},
  \bibinfo{pages}{050402} (\bibinfo{year}{2006}).


\bibitem[{\citenamefont{St\"oferle et~al.}(2006)\citenamefont{St\"oferle,
  Moritz, G\"unter, K\"ohl, and Esslinger}}]{StoeferleEsslinger2006}
\bibinfo{author}{\bibfnamefont{T.}~\bibnamefont{St\"oferle {\em et al.}}},
  \bibinfo{journal}{Phys.~ Rev.~ Lett.} \textbf{\bibinfo{volume}{96}},
  \bibinfo{pages}{030401} (\bibinfo{year}{2006}).

\bibitem[{\citenamefont{Moritz et~al.}(2005)\citenamefont{Moritz, St\"oferle,
  G\"unter, K\"ohl, and Esslinger}}]{MoritzEsslinger2005}
\bibinfo{author}{\bibfnamefont{H.}~\bibnamefont{Moritz {\em et al.}}},
  \bibinfo{journal}{Phys.~ Rev.~ Lett.} \textbf{\bibinfo{volume}{94}},
  \bibinfo{pages}{210401} (\bibinfo{year}{2005}).

\bibitem[{\citenamefont{Jaksch et~al.}(1998)\citenamefont{Jaksch, Bruder,
  Cirac, Gardiner, and Zoller}}]{JakschZoller1998}
\bibinfo{author}{\bibfnamefont{D.}~\bibnamefont{Jaksch {\em et al.}}},
  \bibinfo{journal}{Phys.~ Rev.~ Lett.} \textbf{\bibinfo{volume}{81}},
  \bibinfo{pages}{3108} (\bibinfo{year}{1998}).

\bibitem[{\citenamefont{Gebhard}(1997)}]{Gebhard1997}
\bibinfo{author}{\bibfnamefont{F.}~\bibnamefont{Gebhard}},
  \emph{\bibinfo{title}{The {M}ott Metal-Insulating Transition}}
  (\bibinfo{publisher}{Springer}, \bibinfo{year}{1997}).

\bibitem[{\citenamefont{Essler et~al.}(2005)\citenamefont{Essler, Frahm,
  G\"ohmann, Kl\"umper, and Korepin}}]{EsslerKorepin2005}
\bibinfo{author}{\bibfnamefont{F.}~\bibnamefont{Essler {\em et al.}}},
  \emph{\bibinfo{title}{The One-Dimensional Hubbard Model}}
  (\bibinfo{publisher}{Cambridge University Press}, \bibinfo{year}{2005}).

\bibitem[{\citenamefont{Schulz}(1995)}]{Schulz1995}
\bibinfo{author}{\bibfnamefont{H.~J.} \bibnamefont{Schulz}}, in
  \emph{\bibinfo{booktitle}{Mesoscopic Quantum Physics}}, edited by
  \bibinfo{editor}{\bibfnamefont{E.}~\bibnamefont{Akkermans}}
  (\bibinfo{publisher}{Elsevier (Amsterdam)}, \bibinfo{year}{1995}), vol.
  \bibinfo{volume}{LXI} of \emph{\bibinfo{series}{Les Houches}}.

\bibitem[{\citenamefont{Giamarchi}(2004)}]{Giamarchibook}
\bibinfo{author}{\bibfnamefont{T.}~\bibnamefont{Giamarchi}},
  \emph{\bibinfo{title}{Quantum Physics in One Dimension}}
  (\bibinfo{publisher}{Oxford University Press}, \bibinfo{year}{2004}).

\bibitem[{\citenamefont{Daley et~al.}(2004)\citenamefont{Daley, Kollath,
  Schollw\"ock, and Vidal}}]{DaleyVidal2004}
\bibinfo{author}{\bibfnamefont{A.~J.} \bibnamefont{Daley}},
  \bibinfo{author}{\bibfnamefont{C.}~\bibnamefont{Kollath}},
  \bibinfo{author}{\bibfnamefont{U.}~\bibnamefont{Schollw\"ock}},
  \bibnamefont{and} \bibinfo{author}{\bibfnamefont{G.}~\bibnamefont{Vidal}},
  \bibinfo{journal}{J.~ Stat.~ Mech.: Theor.~ Exp.~}
  \textbf{\bibinfo{volume}{P04005}} (\bibinfo{year}{2004}).

\bibitem[{\citenamefont{White and Feiguin}(2004)}]{WhiteFeiguin2004}
\bibinfo{author}{\bibfnamefont{S.~R.} \bibnamefont{White}} \bibnamefont{and}
  \bibinfo{author}{\bibfnamefont{A.~E.} \bibnamefont{Feiguin}},
  \bibinfo{journal}{Phys.~ Rev.~ Lett.} \textbf{\bibinfo{volume}{93}},
  \bibinfo{pages}{076401} (\bibinfo{year}{2004}).

\bibitem[{\citenamefont{Kollath et~al.}(2005)\citenamefont{Kollath,
  Schollw\"ock, and Zwerger}}]{KollathZwerger2005}
\bibinfo{author}{\bibfnamefont{C.}~\bibnamefont{Kollath}},
  \bibinfo{author}{\bibfnamefont{U.}~\bibnamefont{Schollw\"ock}},
  \bibnamefont{and} \bibinfo{author}{\bibfnamefont{W.}~\bibnamefont{Zwerger}},
  \bibinfo{journal}{Phys.~ Rev.~ Lett.} \textbf{\bibinfo{volume}{95}},
  \bibinfo{pages}{176401} (\bibinfo{year}{2005}).

\bibitem[{\citenamefont{Rigol and Muramatsu}(2004)}]{RigolMuramatsu2004c}
\bibinfo{author}{\bibfnamefont{M.}~\bibnamefont{Rigol}} \bibnamefont{and}
  \bibinfo{author}{\bibfnamefont{A.}~\bibnamefont{Muramatsu}},
  \bibinfo{journal}{Opt.~Commun.} \textbf{\bibinfo{volume}{243}},
  \bibinfo{pages}{33} (\bibinfo{year}{2004}).

\end{thebibliography}


\end{document}